\documentclass[runningheads,a4paper]{llncs}
\usepackage{graphicx, times, amsmath, amssymb, cite, subfigure, flushend, stfloats, bm}
\usepackage[lined,ruled,commentsnumbered]{algorithm2e}
\usepackage{booktabs,multirow}

\usepackage{amssymb}
\usepackage{graphicx}

\usepackage{url}
\urldef{\mailsa}\path|{alfred.hofmann, ursula.barth, ingrid.haas, frank.holzwarth,|
\urldef{\mailsb}\path|anna.kramer, leonie.kunz, christine.reiss, nicole.sator,|
\urldef{\mailsc}\path|erika.siebert-cole, peter.strasser, lncs}@springer.com|
\newcommand{\keywords}[1]{\par\addvspace\baselineskip
\noindent\keywordname\enspace\ignorespaces#1}

\begin{document}

\title{White-Box Target Attack\\ for EEG-Based BCI Regression Problems}
\titlerunning{White-Box Target Attack\\ for EEG-Based BCI Regression Problems}

\author{Lubin~Meng\inst{1} \and Chin-Teng Lin\inst{2}\and Tzyy-Ring Jung\inst{3} \and Dongrui~Wu\inst{1}}
\institute{School of Artificial Intelligence and Automation, \\ Huazhong University of Science and Technology, Wuhan, Hubei, China 
\and
Centre of Artificial Intelligence, Faculty of Engineering and Information Technology, University of Technology, Sydney, Australia 
\and
Swartz Center for Computational Neuroscience, Institute for Neural Computation, \\ University of California San Diego (UCSD), La Jolla, CA 
}

\maketitle

\begin{abstract}
Machine learning has achieved great success in many applications, including electroencephalogram (EEG) based brain-computer interfaces (BCIs). Unfortunately, many machine learning models are vulnerable to adversarial examples, which are crafted by adding deliberately designed perturbations to the original inputs. Many adversarial attack approaches for classification problems have been proposed, but few have considered target adversarial attacks for regression problems. This paper proposes two such approaches. More specifically, we consider white-box target attacks for regression problems, where we know all information about the regression model to be attacked, and want to design small perturbations to change the regression output by a pre-determined amount. Experiments on two BCI regression problems verified that both approaches are effective. Moreover, adversarial examples generated from both approaches are also transferable, which means that we can use adversarial examples generated from one known regression model to attack an unknown regression model, i.e., to perform black-box attacks. To our knowledge, this is the first study on adversarial attacks for EEG-based BCI regression problems, which calls for more attention on the security of BCI systems.
\end{abstract}

\section{Introduction} \label{sect:Intro}

Machine learning has been widely used to solve many difficult tasks. One of them is brain-computer interfaces (BCIs). BCIs enable a user to directly communicate with a computer via brain signals~\cite{BCI}, and have attracted lots of research interest recently~\cite{EEG,Makeig2010}. Electroencephalogram (EEG) is the most frequently used input signal in BCIs, because of its low-cost and non-invasive nature. Three commonly used BCI paradigms are motor imagery (MI)~\cite{Pfurtscheller2001}, event-related potentials (ERP)~\cite{Sutton1965,Wu2018}, and steady-state visual evoked potentials (SSVEP)~\cite{Middendorf2000}. Machine learning can be used to extract more generalizable features~\cite{Zander2011} and construct high-performance models~\cite{Wu2017}, and hence makes BCIs more robust and user-friendly.
	
Recent research has shown that many machine learning models are vulnerable to adversarial examples. By adding deliberately designed perturbations to legitimate data, adversarial examples can cause large changes in the model outputs. The perturbations are usually so small that they are hardly noticeable by a human or a computer program, but can dramatically degrade the model performance. For example, in image recognition, adversarial examples can easily mislead a classifier to give a wrong output~\cite{FGSM}. In speech recognition, adversarial examples can generate audio that sounds meaningless to a human, but be understood as a meaningful voice command by a smart phone~\cite{Carlini2016}. Our recent work \cite{Zhang2019} also showed that adversarial examples can dramatically degrade the classification accuracy of EEG-based BCIs.

There are many different approaches for crafting adversarial examples. Szegedy \emph{et al.}~\cite{Szegedy2014} first discovered the existence of adversarial examples in 2014, and proposed an optimization-based approach, L-BFGS, to find them. Goodfellow \emph{et al.}~\cite{FGSM} proposed a fast gradient sign method (FGSM) in 2014, which can rapidly find adversarial examples by searching for perturbations in the direction the loss has the fastest change. Carlini and Wagner~\cite{CW} proposed the CW method in 2017, which can find adversarial examples with very small distortions.

All above approaches focused on classification problems, which find perturbations that can push the original examples cross the decision boundary. Jagielski \emph{et al.}~\cite{Jagielski2018} conducted the first \emph{non-target} adversarial attacks for linear regression models. This paper considers \emph{target} adversarial attacks for regression problems, which change the model output by a pre-determined amount. Our contributions are:
\begin{enumerate}
\item We propose two approaches, based on optimization and gradient, respectively, to perform white-box target attack for regression problems.
\item We validate the effectiveness of our proposed approaches in two EEG-based BCI regression problems (drowsiness estimation and reaction time estimation). They can craft adversarial EEG trials that a human cannot distinguish from the original EEG trials, but can dramatically change the outputs of the BCI regression model.
\item We show that adversarial examples crafted by our approaches are transferable: adversarial examples crafted from a ridge regression model can also successfully attack a neural network model, and vice versa. This makes black-box attacks possible.
\end{enumerate}
The attacks proposed in this paper may pose serious security and safety problems in real-world BCI applications. For example, an EEG-based BCI system may be used to monitor the driver's drowsiness level and urge him/her to take breaks accordingly. An attack that deliberately changes the estimated drowsiness level from a high value to a low value may overload the driver, and hence cause accidents.

The remainder of this paper is organized as follows: Section~\ref{sect:BK} introduces several typical adversarial attack approaches for classification problems. Section~\ref{sect:reg} proposes two white-box target attack approaches for regression problems. Section~\ref{sect:eval} evaluates the performances of our proposed approaches in two EEG-based BCI regression problems. Section~\ref{sect:conc} draws conclusion.
 	
\section{Adversarial Attacks for Classification Problems} \label{sect:BK}

This section introduces two typical adversarial attack approaches for classification problems, which are extended to regression problems in the next section.

\subsection{Adversarial Attack Types}

Assume a valid benign example $\mathbf{x}\in[0,1]^k$ ($k$ is the dimensionality of $\mathbf{x}$) is classified into Class $y$ by a classifier $f(\mathbf{x})$. It is possible to find an adversarial example $\mathbf{x}'\in[0,1]^k$, which is very similar to the original sample $\mathbf{x}$ according to some distance metric $d$, but is misclassified to $f(\mathbf{x}')\neq{y}$. According to how $f(\mathbf{x}')$ is different from $y$, there can be two types of attacks:
\begin{enumerate}
\item \emph{Target attack}, in which all adversarial examples are classified into a pre-determined class $y'\neq y$.
\item \emph{Non-target attack}, whose goal is to construct adversarial examples that will be misclassified, but does not require them to be misclassified into a particular class.
\end{enumerate}

According to how much knowledge the attacker can obtain about the target model (the model to be attacked), adversarial attacks can also be categorized into:
\begin{enumerate}
\item \emph{White-box attack}, in which the attacker knows all information about the target model, such as its architecture and all parameter values.
\item \emph{Black-box attack}, in which the attacker does not know the architecture and parameters of the target model; instead, he/she can feed some inputs to it and observe its outputs. In this way, he/she can obtain some training examples, and train a substitute model to craft adversarial examples to attack the target model. This approach makes use of the transferability of the adversarial examples~\cite{Papernot2016}.
\end{enumerate}

\subsection{White-Box Target Attack Approaches}
	
This paper considers white-box target attacks only. Assume we know the architecture and all parameters of the classifier $f(\mathbf{x})$. We want to craft an adversarial example $\mathbf{x}'$ from an input $\mathbf{x}$ so that $f(\mathbf{x}')=y_t$, where $y_t$ is a fixed class for all $\mathbf{x}'$.

Two representative target attack approaches for classification problems are:
\begin{enumerate}
%

\item \emph{Carlini and Wagner (CW)}~\cite{CW}, which improves L-BFGS~\cite{Szegedy2014}. It introduces a new variable $\boldsymbol{\omega}$ so that
    \begin{align}
     \boldsymbol{\delta}=\frac{1}{2}(\tanh(\boldsymbol{\omega})+1) - \mathbf{x} \label{eq:delta}
     \end{align}
      automatically satisfies the constraint $\mathbf{x}'=\mathbf{x}+\boldsymbol{\delta}=\frac{1}{2}(\tanh(\boldsymbol{\omega})+1)\in[0,1]^k$.  $\boldsymbol{\omega}$ in (\ref{eq:delta}) is the variable to be optimized, which can assume any value in $(-\infty,\infty)$.

      Given $\mathbf{x}$, $\boldsymbol{w}$ is found through:
\begin{align}
&\min \left[\left\rVert \frac{1}{2}(\tanh(\boldsymbol{\omega})+1)-\mathbf{x} \right\lVert_{2}
+ c \cdot \ell\left(\frac{1}{2}(\tanh(\boldsymbol{\omega})+1)\right)\right]\nonumber\\
=&\min \left[\rVert \boldsymbol{\delta} \lVert_{2} + c \cdot \ell(\mathbf{x}')\right]
\end{align}			
where $c$ is a trade-off parameter, and
\begin{align}
\ell(\mathbf{x}') = \max\left(\max_{i\neq{y_t}}Z(\mathbf{x}')_i -Z(\mathbf{x}')_{y_t}, -\lambda\right),
\end{align}
in which $Z(\mathbf{x}')_i$ is the logits of the target model in Class~$i$, and $\lambda$ controls the confidence of the adversarial example. A large $\lambda$ forces the adversarial example to be classified into the target class $y_t$ with high confidence.

\item \emph{Iterative Target Class Method (ITCM)}\footnote{It is also called the iterative least-likely class method in \cite{Kurakin2017}.} \cite{Kurakin2017}, which modifies FGSM~\cite{FGSM}, an efficient approach for non-target attacks:
\begin{align}
\mathbf{x}' = \mathbf{x} + \epsilon \cdot \mathrm{sign}(\nabla_{\mathbf{x}}J(\mathbf{x}, y_{true})), \label{eq:FGSM}
\end{align}
where $\epsilon$ controls the amplitude of the perturbation, $J$ is a loss function, and $y_{true}$ is the true label of $\mathbf{x}$.

ITCM performs target attack by replacing $y_{true}$ in (\ref{eq:FGSM}) by the target class $y_t$. It also improves the attack performance by taking multiple small steps of $\alpha$ in the gradient direction and clipping the maximum perturbation to $\epsilon$, instead of taking a single large step of $\epsilon$ in (\ref{eq:FGSM}):
\begin{align}
\mathbf{x}'_{0}&=\mathbf{x}, \\
\mathbf{x}'_{m+1} &= \mathrm{Clip}_{\mathbf{x}, \epsilon}\{\mathbf{x}'_m-\alpha \cdot \mathrm{sign}(\nabla_{\mathbf{x}'_m}J(\mathbf{x}'_m,y_{t})) \}, \label{eq:clip}
\end{align}
where $\mathrm{Clip}_{\mathbf{x}, \epsilon}(\mathbf{x}')$ ensures the difference between each dimension of $\mathbf{x}'$ and the corresponding dimension of $\mathbf{x}$ does not exceed $\epsilon$.
\end{enumerate}

\section{White-Box Target Attack for Regression Problems} \label{sect:reg}

The section proposes two white-box target attack approaches for regression problems.

Let $\mathbf{x}$ be an input, $y$ the groundtruth output, and $g(\mathbf{x})$ the regression model. Target attack aims to generate a small perturbation $\boldsymbol{\delta}$ such that the adversarial example $\mathbf{x}' = \mathbf{x} + \boldsymbol{\delta}$ can change the regression output to $g(\mathbf{x}') \geq y+t$, where $t>0$ is a predefined target\footnote{The regression output can also be changed to $g(\mathbf{x}') \leq y-t$. Without loss of generality, $g(\mathbf{x}') \geq y+t$ is considered in this paper.}:
\begin{align}
\min_{\mathbf{x}'} \quad \lVert \mathbf{x}'-\mathbf{x} \rVert_{2}, \quad \mathrm{s.t.} \quad g(\mathbf{x}') - y \geq t
\end{align}

\subsection{CW for Regression (CW-R)}

To extend the CW target attack approach from classification to regression, we optimize the following loss function:
\begin{align}
\min_{\boldsymbol{\omega}} \left[ \left\lVert \frac{1}{2}(\tanh(\boldsymbol{\omega})+1) -\mathbf{x} \right\rVert_{2} + c \cdot \ell(\mathbf{x}, \boldsymbol{\omega}, t)\right], \label{eq:CWP}
\end{align}
where
\begin{align}
\ell(\mathbf{x}, \boldsymbol{\omega}, t)
 =& \max\left\{g(\mathbf{x}) + t -g\left(\mathbf{x}+\frac{1}{2}(\tanh(\boldsymbol{\omega})+1)\right),0\right\}
 \label{eq:ellCW}\\
 =&\max\{g(\mathbf{x})+t-g(\mathbf{x}'),0\}.
\end{align}
The constructed adversarial example is then $\mathbf{x}'=\mathbf{x}+\frac{1}{2}(\tanh(\boldsymbol{\omega})+1)$.
		
The pseudocode of the proposed CW method for regression (CW-R) is shown in Algorithm~1. It uses iterative binary search to find the optimal trade-off parameter $c$.

\begin{algorithm}
\KwIn{$\mathbf{x}$, the original example\;
\hspace*{10mm}$g(\mathbf{x})$, the target regression model\;
\hspace*{10mm}$t$, the minimum change of the output\;
\hspace*{10mm}$M$, the number of the iterations\;
\hspace*{10mm}$c_{0}$, initialization of the trade-off parameter\;
\hspace*{10mm}$N$, the number of binary search steps for the optimal trade-off parameter.}
\KwOut{$\mathbf{x}'$, the adversarial example.}
Initialize $ c \leftarrow c_{0}$, $\boldsymbol{\omega}_1 \leftarrow random$, $d_{\min} \leftarrow \infty$, $\overline{c} \leftarrow 1e4$, $\underline{c} \leftarrow 0$\\

\For{$n=1:N$}
{
\For{$m=1:M$}
{
$\mathbf{x}'_m \leftarrow \frac{1}{2}(\tanh(\boldsymbol{\omega}_m)+1)$\;
$\ell \leftarrow \lVert \mathbf{x}'_m - \mathbf{x} \rVert_{2} + c\cdot \max\{g(\mathbf{x})+t-g(\mathbf{x}'_m), 0\}$\;
$\boldsymbol{\omega}_{m+1} \leftarrow \boldsymbol{\omega}_m - \alpha \cdot \frac{\partial \ell} {\partial \boldsymbol{\omega}_m}$\;
\uIf{$g(\mathbf{x}'_{m}) \geq g(\mathbf{x})+t$  {\bf and} $\lVert \mathbf{x}'_m - \mathbf{x} \rVert_{2} \leq d_{\min}$}
{
$\mathbf{x}' \leftarrow \mathbf{x}'_m$\;
$d_{\min} \leftarrow \lVert \mathbf{x}'_m - \mathbf{x} \rVert_{2}$\;}}
\tcp{Update $c$ using binary search}
$c \leftarrow (\overline{c} + \underline{c})/2$\;
\uIf{$g(\mathbf{x}') \geq g(\mathbf{x}) + t$}
{
$\overline{c} \leftarrow c$\;}
\Else
{

$\underline{c} \leftarrow c$\;}}
\KwRet $\mathbf{x}'$
\caption{CW for regression (CW-R). }\label{alg:CW}
\end{algorithm}
		
\subsection{Iterative Fast Gradient Sign Method for Regression (IFGSM-R)}

Iterative fast gradient sign method for regression (IFGSM-R) extends ITCM from classification to regression.

Define the loss function
\begin{align}
\ell(\mathbf{x},\mathbf{x}_m',t)=\max\{g(\mathbf{x})+t-g(\mathbf{x}_m'),0\}, \label{eq:ellFGSMP}
\end{align}
which is essentially the same as (\ref{eq:ellCW}), except that a change of variable is not used here. Then, the adversarial example can be iteratively calculated as:
\begin{align}
\mathbf{x}'_{0} &= \mathbf{x}, \\
\mathbf{x}'_{m+1} &= \mathrm{Clip}_{\mathbf{x},\epsilon}\{\mathbf{x}'_m - \alpha \cdot \mathrm{sign}(\nabla_{\mathbf{x}'_m}\ell(\mathbf{x},\mathbf{x}'_m,t))\}. \label{eq:xn1}
\end{align}

The pseudocode of the proposed IFGSM-R is shown in Algorithm~2.

\begin{algorithm}
\KwIn{$\mathbf{x}$, the original example\;
\hspace*{10mm}$g(\mathbf{x})$, the target regression model\;
\hspace*{10mm}$t$, the minimum change of the output\;
\hspace*{10mm}$M$, the number of iterations\;
\hspace*{10mm}$\epsilon$, the upper bound of the perturbation\;
\hspace*{10mm}$\alpha$, the step size.}
\KwOut{$\mathbf{x}'$, the adversarial example.}
$ \mathbf{x}'_{0} = \mathbf{x} $\;
\For{$m=0:M$}
{
$\mathbf{x}'_{m+1} \leftarrow \mathrm{Clip}_{\mathbf{x}, \epsilon}(\mathbf{x}'_m - \alpha \cdot \mathrm{sign}(\nabla_{\mathbf{x}'_m}\ell(\mathbf{x},\mathbf{x}'_m, t)))$, using $\ell(\mathbf{x},\mathbf{x}_m',t)$ in (\ref{eq:ellFGSMP})\;
}
\KwRet $\mathbf{x}_{M+1}'$
\caption{Iterative fast gradient sign method for regression (IFGSM-R). }\label{alg:ifgsm}
\end{algorithm}

\section{Experiments and Results} \label{sect:eval}

This section evaluates the performances of the two proposed white-box target attack approaches in two BCI regression problems.

\subsection{The Two BCI Regression Problems}

We used the following two BCI regression datasets in our experiments:
\begin{enumerate}
\item \emph{Driving}. The driving dataset was collected from 16 subjects (ten males, six females; age 24.2 $\pm$ 3.7), who participated in a sustained-attention driving experiment~\cite{Chuang2014,Wu2017}. Our task was to predict the drowsiness index from the EEG signals, which were recorded using 32 channels with a sampling rate of 500 Hz. Our preprocessing and feature extraction procedures were identical to those in \cite{Wu2015}. We applied a [1,50] Hz band-Rass filter to remove artifacts and noise, and then downsampled the EEG signals from 500 Hz to 250 Hz. Next, we computed the average power spectral density in the theta band (4-7 Hz) and alpha band (7-13 Hz) for each channel, and used them as our features, after removing abnormal channels. Since data from one subject were not recorded correctly, we only used 15 subjects in our paper. Each subject had about 1000 samples. More details about this dataset can be found in \cite{Chuang2014,Wu2015}.

\item \emph{PVT}. A psychomotor vigilance task (PVT) \cite{Dinges1985} uses reaction time (RT) to measure a subject's response speed to a visual stimulus. Our dataset \cite{Wu2017a} consisted of 17 subjects (13 males, four females; age 22.4 $\pm$ 1.6), each with 465-843 trials. The 64-channel EEG signals were preprocessed using the standardized early-stage EEG processing pipeline (PREP)~\cite{Bigdely-Shamlo2015}. Then, they were downsampled from 1000 Hz to 256 Hz, and passed through a $[1,20]$ Hz band-Rass filter. Similar to the driving dataset, we also computed the average power spectral density in the theta band (4-7 Hz) and alpha band (7-13 Hz) for each channel as our features. The goal was to predict a user's RT from the EEG signals. More details about this dataset can be found in \cite{Wu2017a,drwuSF2018}.
\end{enumerate}

\subsection{Experimental Settings and Performance Measures}

We performed white-box target attack on the two BCI regression datasets. Assume the attacker knows all information about the regression model, i.e., its architecture and parameters. We crafted adversarial examples that can change the regression model output by a pre-determined amount.

Two regression models were considered. The first was ridge regression (RR) with ridge parameter $0.1$. The second was a multi-layer perceptron (MLP) neural network with two hidden layers and 50 nodes in each layer. We used the Adam optimizer~\cite{Adam} and the root mean squared error (RMSE) as the loss function. Early stopping was used to reduce over-fitting.

Two attack scenarios were considered:
\begin{enumerate}
\item \emph{Within-subject attack}. For each individual subject, we randomly chose $90\%$ data for training the RR model and the rest $10\%$ for testing. For the MLP, we further randomly set apart $10\%$ of the training set as the validation set in early stopping. We computed the test RMSE for each subject, and also their average across all subjects.
\item \emph{Cross-subject attack}. Each time we picked one subject as the test subject, and concatenated data from all remaining subjects together to train the RR model. For the MLP, $90\%$ of these data were randomly selected for training, and the remaining $10\%$ for validation in early stopping. RMSEs were computed on the test subject.
\end{enumerate}

Attack success rate (ASR) and distortion were used to evaluate the attack performance. The ASR was defined as the percentage of adversarial examples whose prediction satisfied $g(\mathbf{x}') \geq g(\mathbf{x}) + t$, where $t>0$ was our targeted change. The distortion was computed as the $L_{2}$ distance between the adversarial example and the original example.

\subsection{Experimental Results}

The baseline regression performances on the original (unperturbed) EEG data are shown in the first panel of Table~\ref{tab:perf}, where ``mean output (MO)" is the mean of the regression outputs for all EEG trials. For each regression model on each dataset, the cross-subject RMSE was always larger than the corresponding within-subject RMSE, which is intuitive, because individual differences make it difficult to develop a model that generalizes well across subjects.

\begin{table*}  \center
\caption{Baseline regression performances on the original EEG data, and the attack performances by CW-R, IFGSM-R, and random noise. $t=0.2$ was used.} \setlength{\tabcolsep}{0.5mm} \label{tab:perf}
\begin{tabular}{c|c|cc|cc|cc|cc}\toprule
\multicolumn{2}{c}{Scenario} &\multicolumn{4}{|c}{Within-subject}  &\multicolumn{4}{|c}{Cross-subject} \\ \cline{1-10}
\multicolumn{2}{c}{Dataset} &\multicolumn{2}{|c}{Driving} &\multicolumn{2}{|c}{PVT}	&\multicolumn{2}{|c}{Driving} &\multicolumn{2}{|c}{PVT}	\\ \cline{1-10}
\multicolumn{2}{c|}{Model} &RR &MLP	&RR &MLP	 &RR &MLP &RR &MLP \\ \midrule	
\multirow{2}{*}{Baseline}
&RMSE &$.1766$ &$.1355$ &$.1293$ &$.1445$ &$.2207$ &$.2124$ &$.2255$ &$.2433$ \\
&MO   &$.3805$ &$.3715$ &$.5262$ &$.5318$ &$.2499$ &$.2371$ &$.5384$ &$.5333$ \\ \midrule
\multirow{4}{*}{CW-R}
&RMSE &$.2732$ &$.2368$ &$.2569$ &$.2693$ &$.2976$ &$.2753$ &$.3349$ &$.3405$ \\
&MO   &$.5805$ &$.5717$ &$.7262$ &$.7319$ &$.4499$ &$.4374$ &$.7385$ &$.7337$ \\
&ASR  &$99.59\%$ &$99.94\%$ &$99.68\%$ &$100\%$ &$99.97\%$ &$100\%$ &$99.81\%$ &$99.91\%$ \\
&Distortion &$2.5835$ &$.5858$ &$6.7537$ &$3.7048$ &$.8687$ &$.4008$ &$.4678$ &$.5333$ \\\midrule
\multirow{4}{*}{IFGSM-R}
&RMSE &$.2858$ &$.2788$ &$.2719$ &$.2828$ &$.3176$ &$.3193$ &$.3553$ &$.3857$ \\
&MO   &$.5967$ &$.6168$ &$.7478$ &$.7515$ &$.4790$ &$.4967$ &$.7657$ &$.7884$ \\
&ASR  &$96.88\%$ &$99.54\%$ &$92.29\%$ &$99.19\%$ &$98.97\%$ &$99.97\%$ &$99.94\%$ &$99.56\%$ \\
&Distortion &$6.9672$ &$1.7338$ &$14.3129$ &$7.8370$ &$2.6852$ &$1.3415$ &$1.5272$ &$2.3071$ \\\midrule
\multirow{4}{*}{Random Noise}
&RMSE &$.1766$ &$.1355$ &$.1293$ &$.1445$ &$.2207$ &$.2124$ &$.2255$ &$.2433$ \\
&MO   &$.3805$ &$.3715$ &$.5262$ &$.5318$ &$.2499$ &$.2371$ &$.5384$ &$.5333$ \\
&ASR  &$0.00\%$ &$0.00\%$ &$0.00\%$ &$0.00\%$ &$0.00\%$ &$0.00\%$ &$0.00\%$ &$0.00\%$ \\
&Distortion &$7.1023$ &$1.8663$ &$14.5002$ &$8.0664$ &$2.8018$ &$1.4866$ &$1.6348$ &$2.4029$ \\\midrule
\end{tabular}
\end{table*}

We set $t=0.2$ in both CW-R and IFGSM-R, and called the attack a success if $g(\mathbf{x}') \geq g(\mathbf{x}) + t$. $N=9$ and $c_0=0.01$ were used in CW-R. $M=25$, $\alpha=0.001$, and grid search for $\epsilon\in\{0.001,0.002,...,0.03\}$ were used in IFGSM-R. We used $L_{2}$ distance to measure the distortion of the adversarial examples. The attack performances are shown in the second and third panels of Table~\ref{tab:perf}:
\begin{enumerate}
\item The RMSEs after CW-R and IFGSM-R attacks were always much larger than those before the attacks, indicating that the attacks dramatically changed the characteristics of the model output.
\item For each regression model on each dataset, the mean output of the adversarial examples was always larger than that of the original examples by at least $t$, which was our target. This suggests that both CW-R and IFGSM-R were effective.
\item The ASRs of both CW-R and IFGSM-R were always close to 100\%, indicating that almost all attacks were successful. A closer look revealed that the ASR of CW-R was always slightly larger than the corresponding ASR of IFGSM-R, and the RMSE, mean output, and distortion of CW-R were always smaller than the corresponding quantities of IFGSM-R, i.e., CW-R was generally more effective than IFGSM-R. However, the computational cost of CW-R was much higher than IFGSM-R.
\end{enumerate}

It's also interesting to check if adding random noise can significantly degrade the regression performance; if so, then no deliberate adversarial example crafting is needed. To this end, we performed attacks by adding random Gaussian noise $\mathcal{N}(0,\sigma)$ to the original examples, where $\sigma$ was chosen so that the resulted distortion approximately equaled the maximum distortion introduced by CW-R and IFGSM-R. The corresponding attack performances are shown in the last panel of Table~\ref{tab:perf}. Though the distortion was large, random Gaussian noise almost did not change the regression RMSE and the mean output, and its ASR was always $0.00\%$, suggesting that sophisticated attack approaches like CW-R and IFGSM-R are indeed needed.	

%

\begin{figure}[htbp] \centering
\subfigure[]{\includegraphics[width=.48\linewidth,clip]{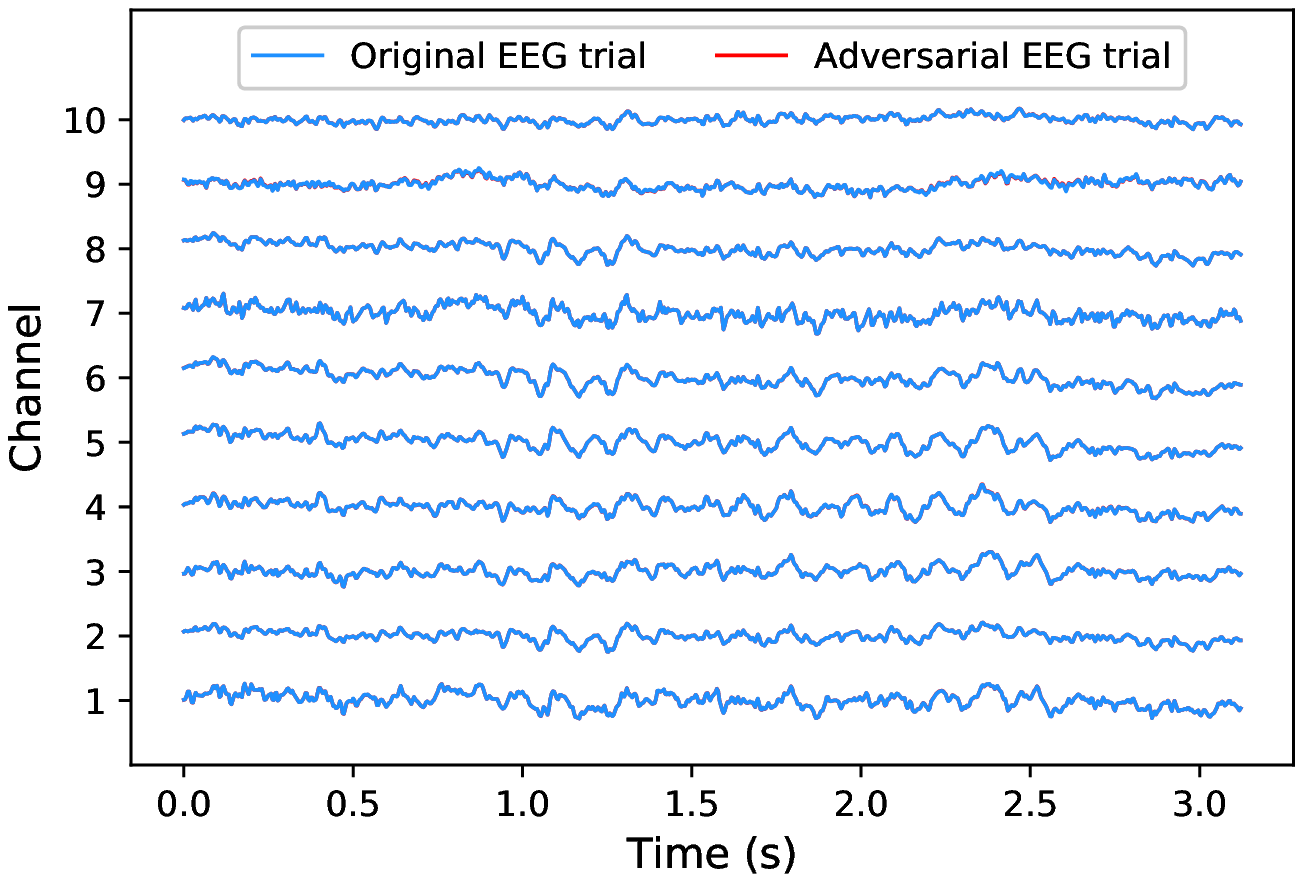}\label{fig:driving_attack}}
\subfigure[]{\includegraphics[width=.48\linewidth,clip]{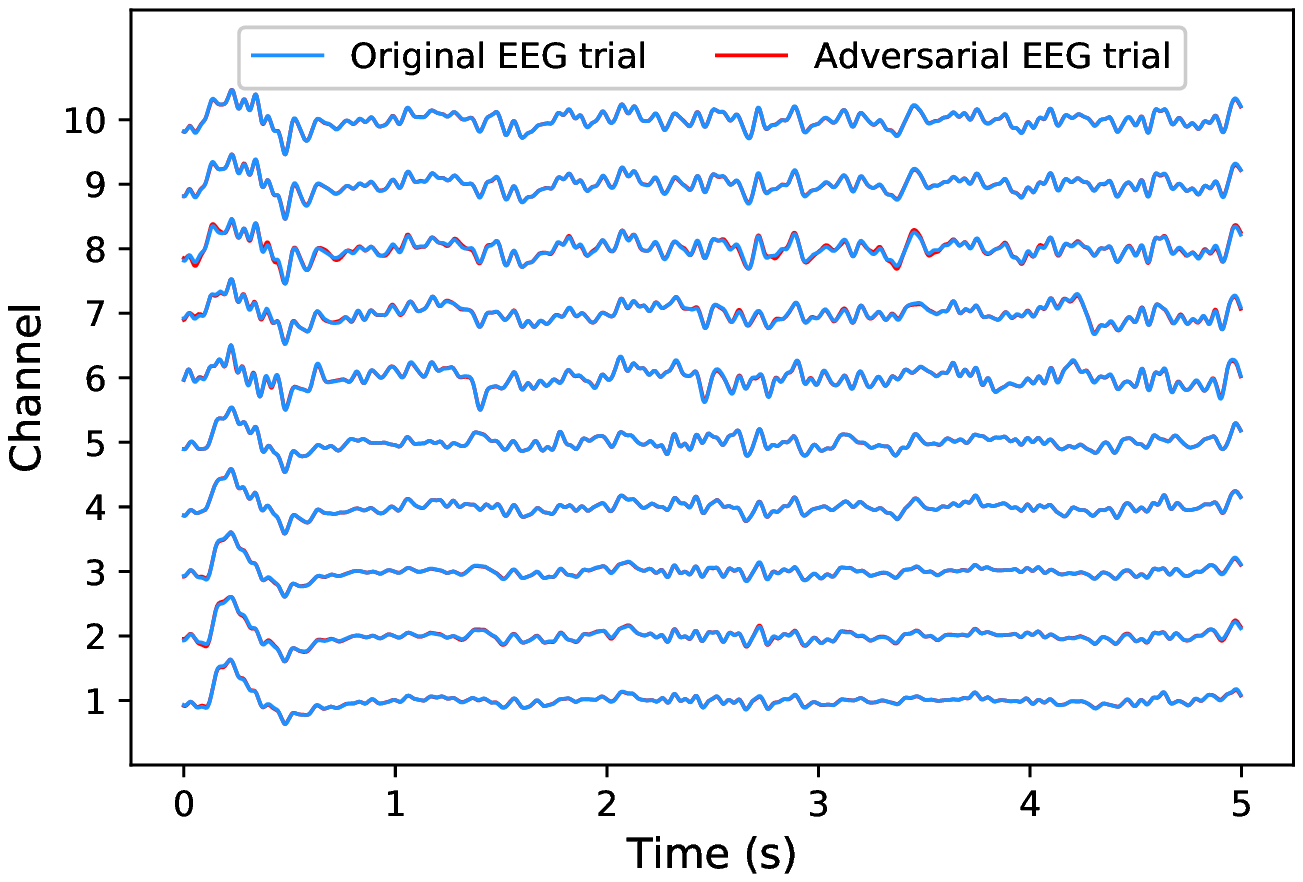}\label{fig:pvt_attack}}
\caption{Examples of the original EEG trials (blue) and the adversarial trials (red), generated by CW-R. (a) driving; (b) PVT. The blue and red curves almost completely overlap.} \label{fig:attack}
\end{figure}

Some examples of the original EEG trials and those after adding adversarial perturbations are shown in Fig.~\ref{fig:attack}. The differences between the original and adversarial trials were too small to be distinguished by a human, which should also be very difficult to be detected by a computer algorithm.

\subsection{Spectrogram Analysis}

\begin{figure*}[htpb] \centering
\subfigure[]{\includegraphics[width=\linewidth,clip]{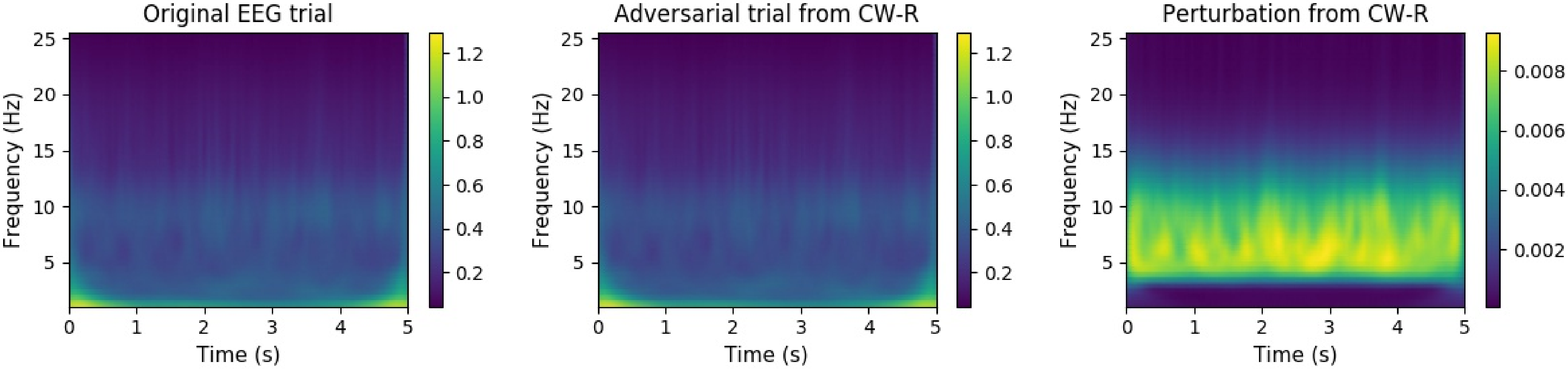}\label{fig:a_cw}}
\subfigure[]{\includegraphics[width=\linewidth,clip]{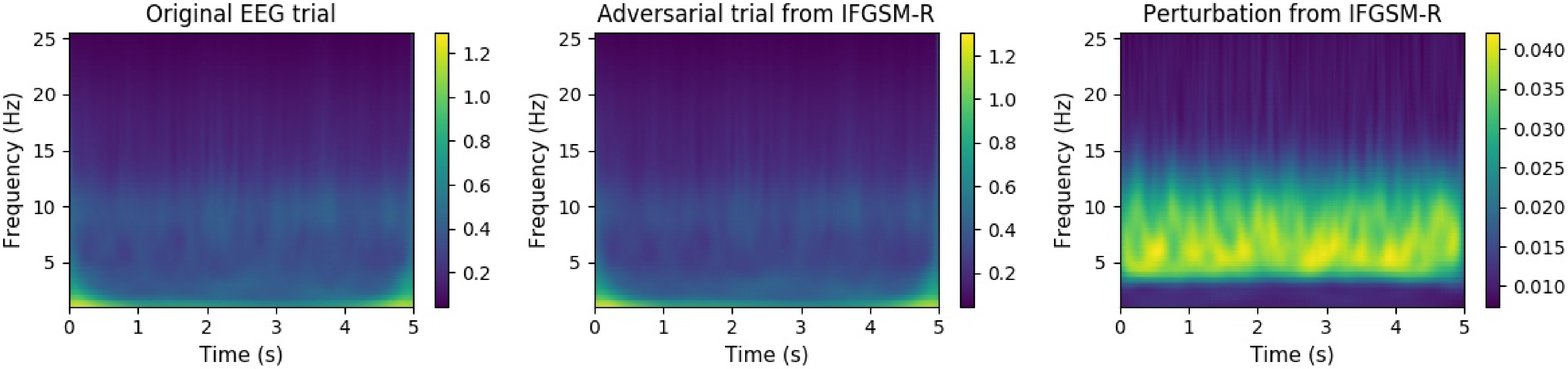}\label{fig:a_ifgsm}}
\caption{Mean spectrogram of all original EEG trials (first column), mean spectrogram of all successful adversarial examples (second column), and mean spectrogram of the perturbations (third column), from MLP on the PVT dataset. Channel $C_z$ was used. (a) CW-R; (b) IFGSM-R.} \label{fig:SA}
\end{figure*}

This section utilizes spectrogram analysis to further understand the characteristics of the adversarial examples. We computed the mean spectrogram of all EEG trials, the mean spectrogram of all successful adversarial examples, and the mean spectrogram of the corresponding perturbations, using wavelet decomposition. Fig.~\ref{fig:SA} shows the results, where the adversarial examples were designed for MLP on the PVT dataset. There is no noticeable difference between the mean spectrograms of the original EEG trials and the adversarial examples crafted by our two approaches. This suggests that adversarial examples are difficult to distinguish from spectrogram analysis.

The third column of Fig.~\ref{fig:SA} shows the difference between the mean spectrograms in the first two columns. Note that the amplitudes were much smaller than those in the first two columns. The patterns of those two perturbations are similar. The energy of those perturbations was concentrated in [3,10] Hz, and was almost uniformly distributed in the entire time domain.
    	
\subsection{Transferability of Adversarial Examples between Different Regression Models}

The transferability of adversarial examples means that adversarial examples designed to attack one model may also be used to attack a different model. This property makes black-box attacks possible, where we have no information about the target regression model at all \cite{Papernot2016,Papernot2017}.

\begin{figure*}[htpb] \centering
\subfigure[]{\includegraphics[width=.48\linewidth,clip]{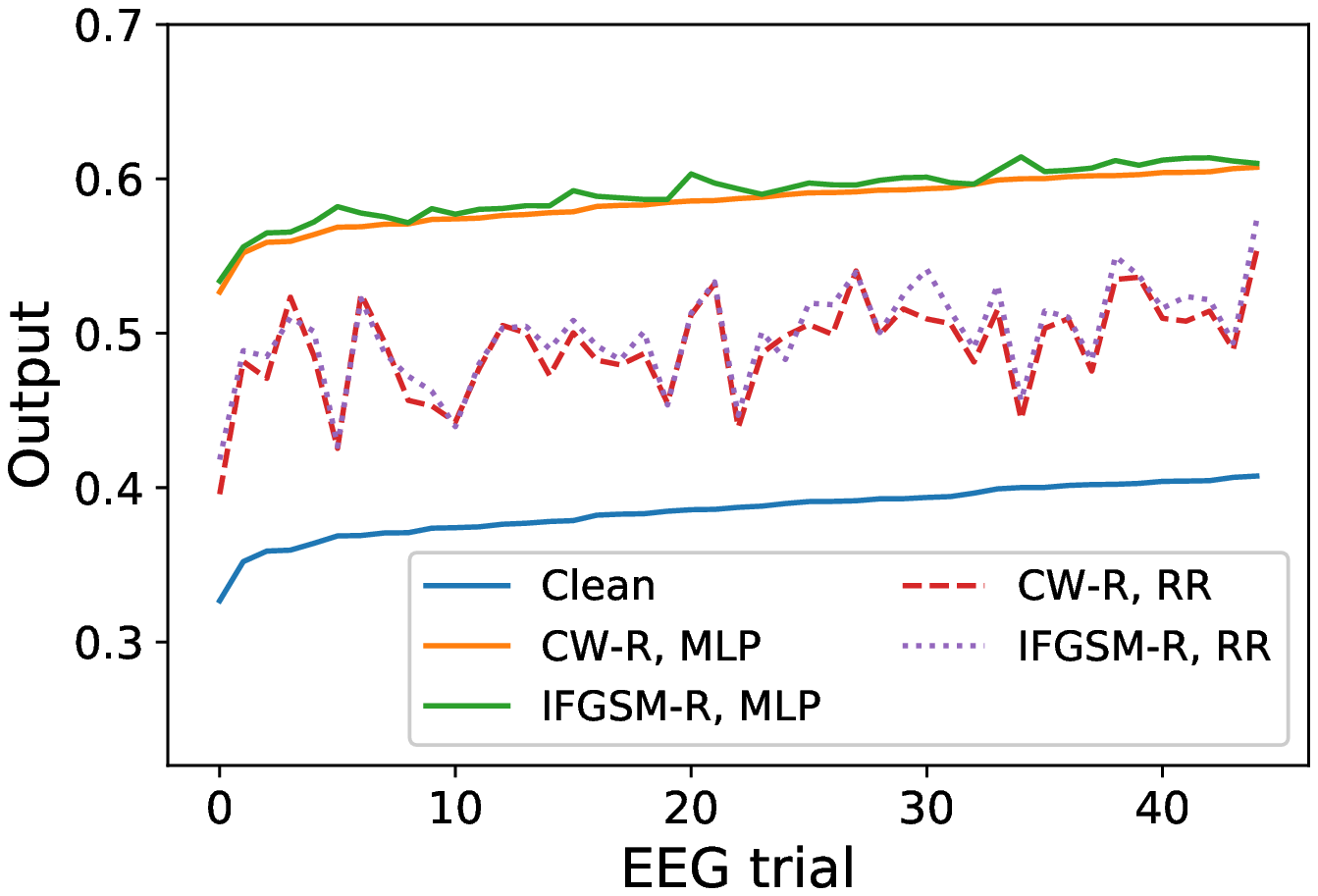}\label{fig:dnn2ridge}}
\subfigure[]{\includegraphics[width=.48\linewidth,clip]{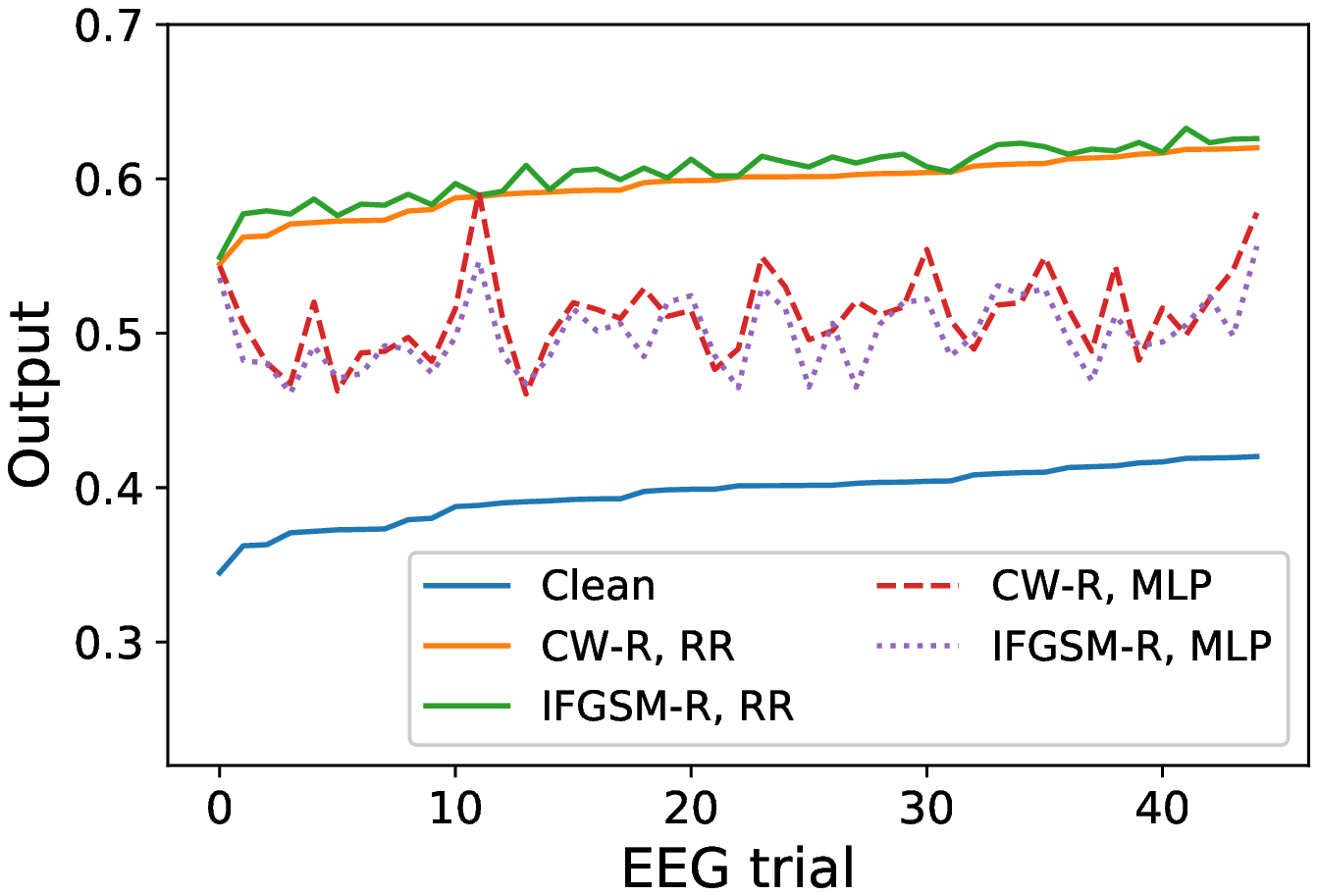}\label{fig:ridge2dnn}}
\caption{Outputs of the adversarial examples, when transferred from (a) MLP to RR, and (b) RR to MLP, in within-subject attacks on the PVT dataset.} \label{fig:transferability}
\end{figure*}

Fig.~\ref{fig:transferability} shows the mean output, when adversarial examples designed from MLP were used to attack the RR model [Fig.~\ref{fig:dnn2ridge}], and vice versa [Fig.~\ref{fig:ridge2dnn}], in within-subject attacks on the PVT dataset. In Fig.~\ref{fig:dnn2ridge}, although the attack performance on RR degraded compared with the attack performance on MLP, the adversarial examples still dramatically changed the outputs of RR. Fig.~\ref{fig:ridge2dnn} is similar. These demonstrate that adversarial examples generated by CW-R and IFGSM-R are also transferrable, and hence may be used in black-box attacks.
    	
\section{Conclusions} \label{sect:conc}

This paper has proposed two white-box target attack approaches, CW-R and IFGSM-R, for regression problems, and applied them to two EEG-based BCI regression problems. Both approaches can successfully change the model output by a pre-determined amount. Generally, CW-R achieved better attack performance than IFGSM-R, in terms of a larger ASR and a smaller distortion; however, its computational cost is higher than IFGSM-R. We also verified that the adversarial examples crafted from both CW-R and IFGSM-R are transferrable, and hence adversarial examples generated from a known regression model can also be used to attack an unknown regression model, i.e., to perform black-box attacks.

To our knowledge, this is the first study on adversarial attacks for EEG-based BCI regression problems, which calls for more attention on the security of BCI systems. Our future research will study how to defend such attacks.

\end{document}